\documentclass[12pt,preprint]{aastex}

\begin{document}

\title{A possible common halo of the Magellanic Clouds}

\author{Kenji Bekki} 
\affil{
School of Physics, University of New South Wales, Sydney 2052, Australia}

\begin{abstract}

Recent observational and theoretical studies on the three-dimensional
(3D) space motions of the Large and the Small Magellanic Clouds 
(LMC and SMC, respectively) have strongly suggested  that
the latest proper motion measurements of the Magellanic Clouds (MCs)
are consistent with their  orbital evolution models  in which
the MCs have arrived 
in the Galaxy quite recently for the first time.
The suggested orbital models appear to be seriously
inconsistent with the tidal interaction models
in which 
the Magellanic Stream (MS)
can be formed as a result of the mutual tidal interaction
between the MCs and the Galaxy
for the last $\sim 2$ Gyr.
Based on orbital models of the MCs,
we propose  that if the MCs have a common diffuse dark halo with the mass
larger than $\sim 2 \times 10^{10} M_{\odot}$,
the MCs can not only have the present 3D 
velocities consistent with the latest proper motion
measurements but also interact strongly  with each other and
with the Galaxy
for the last 2 Gyr.
These results imply that  
if the observed proper motions of the MCs are true ones of the centers
of mass for  the MCs,
the common halo of the MCs would need to be considered in
constructing  self-consistent MS formation models.
We discuss whether the origin of the possible common halo
can be closely associated  either with the past binary formation
or with the MCs having been in a small group.

\end{abstract}


\section{Introduction}

Recent proper motion measurements of the MCs by the
Advanced Camera for Surveys (ACS) on the  {\it Hubble
Space Telescope (HST)} have reported that
the LMC and the SMC have significantly high Galactic
tangential velocities ($367 \pm 18$ km s$^{-1}$ 
and $301 \pm 52$ km s$^{-1}$, respectively)
and thus suggested that the MCs could be unbound
from each other (Kallivayalil et al. 2006; K06).
Piatek et al. (2008)
have independently  analyzed  the same data sets
as those used by K06 and confirmed the proper motions
of the MCs derived from  K06.
Besla et al. (2007, B07)  extensively investigated the long-term orbital
evolution of the MCs by using the results of K06
and thereby suggested  that the MCs  have recently
arrived in the Galaxy for the first time (``the first passage scenario'').

These observational and theoretical results on the 3D space motions
of the MCs appears to be seriously inconsistent with
the tidal interaction models in which the MS can be formed
as a result of the strong tidal interaction between the MCs
and the Galaxy (e.g., Gardiner \& Noguchi 1996, GN):
the MCs are required not only to keep their binary status 
but also to interact strongly with the Galaxy
at least for the
last $\sim 2$ Gyr
in the  models.
Recently Bekki \& Chiba (2008) have shown that the tidal interaction
models consistent with the results by K06 can not reproduce
well the observed  location of the MS on the sky, though
they did not investigate all possible orbits  consistent
with K06.
Given that the tidal interaction models of the MS formation
can explain not only the fundamental properties of the MS
but also the presence of the leading arm features (GN),
it is worth while to discuss whether the tidal interaction 
models consistent with K06 can be constructed by considering
some new physical processes 
that have not been so far
included in the previous models of the MC evolution.

The purpose of this Letter is to propose
that if the MCs have a common diffuse dark halo,
the halo can play an important  role in 
the long-term orbital evolution  of the MCs.
Based on orbital models of the MCs with and without common halos,
we investigate whether the MCs can keep their binary status
at least for the last 2 Gyr 
for their  present  3D velocities consistent with K06.
We demonstrate that some models with common halos
can keep the binary status of the MCs and
thus show strong tidal interaction between the LMC and the SMC
and between MCs and the Galaxy for the last 2 Gyr.
This work is inspired by recent numerical simulations  which have clearly shown
the concurrent  accretion of multiple satellite systems 
onto galaxy-scale halos in a hierarchical galaxy formation scenario
(e.g., Sales et al. 2007; Li \& Helmi 2008; Ludlow et al. 2008).

\section{The model}

\subsection{The common halo scenario}

We investigate orbital evolution of the MCs
with respect to the Galaxy  within the last $\sim 2.7$ Gyr 
for a given set of initial parameters (e.g., velocities
and masses of the MCs)
by using a backward integration
scheme (GN).
Since the present orbital model for the MCs is
almost the same as those adopted in previous studies
(GN; Bekki \& Chiba 2005, BC05),
we briefly describe the model in the present study.
A key difference between the present and the previous
models is that the orbits of the MCs are influenced
by the hypothesized common halo 
that is sinking into the inner region of the Galaxy
via dynamical friction 
in the present study.
We adopt the present velocities of the MCs by K06
for {\it all models}  so that we can discuss whether
we can construct orbital evolution models
in which (i) the present velocities are consistent
with K06 and (ii) the MCs can keep their binary status
for the last 2 Gyr.

The gravitational potential of the Galaxy
is assumed to have the logarithmic potential (i.e., the isothermal
density distribution)
with the constant rotational velocity of 220 km s$^{-1}$.
The LMC is assumed to have the Plummer potential;
\begin{equation}
 {\Phi}_{\rm L}(r_{\rm L})=-M_{\rm L}/{({r_{\rm L}}^2+a_{\rm
L}^2)}^{0.5},
\end{equation}
where $M_{\rm L}$, $r_{\rm L}$, and $a_{\rm L}$ are
the total mass of the LMC, the distance from the LMC, and
the effective radius, respectively. 
The SMC is also assumed to have the 
Plummer potential with the mass of $M_{\rm S}$
and the effective radius of   $a_{\rm S}$.
We adopt the same values
of  $a_{\rm L}$ (=3 kpc) 
and  $a_{\rm S}$ (=2 kpc)  as previous numerical studies
adopted (e.g., GN). 
Although we investigated the models
with $10^{10} M_{\odot} \le M_{\rm L} \le 2 \times 10^{10} M_{\odot}$
and with $3 \times 10^{9} M_{\odot} \le M_{\rm S} \le 6 \times 10^{9} M_{\odot}$,
we mainly show the results of the models with $ M_{\rm L} =  2 \times 10^{10}
M_{\odot}$ and $ M_{\rm S} =  3 \times 10^{9} M_{\odot}$,
which were extensively discussed in previous models (GN, BC05).

The common halo is assumed to have a mass of  $ M_{\rm ch}$
and the  Plummer potential with the effective radius of $a_{\rm ch}$.
This mass of the halo does not include the  masses of 
individual halos in the MCs.
We consider the dynamical friction due to the presence of the Galactic
dark matter halo  for the common halo 
and adopt the following expression (Binney \& Tremaine 1987);
\begin{equation}
F_{\rm fric, G}=-0.428\ln {\Lambda}_{\rm G} \frac{GM^2}{r^2},
\end{equation}
where $r$ is the distance of the common halo
from the center of the Galaxy and $M=M_{\rm ch}$ for the halo.
We adopt the reasonable value of 3.0
for the Coulomb logarithm ${\Lambda}_{\rm G}$
(GN). Although we investigated models with different initial 3D velocities
of the common halo,
we only show the results of the ``successful models'' in which
the common halo can play a role in keeping the binary status
of the MCs for the last 2 Gyr. The initial positions ($x$,$y$,$z$)
and velocities ($v_{\rm x}$,$v_{\rm y}$,$v_{\rm z}$)
with respect to the Galatic center
for the MCs and the common halo are summarized in the Table 1.

The  present velocities of the common halo of the MCs
adopted in the successful models
correspond to those for the center of mass of the LMC and the SMC
used in GN: the total velocity ($|v|$) of the common halo
is 291 km s$^{-1}$, which is significantly smaller than
$|v|=378$ km s$^{-1}$ derived by K06 for the LMC.
We choose these values so that we can show how the common halo
controls the orbital evolution of the MCs, 
{\it if the common halo has a significantly smaller
$|v|$ than the LMC.}
For convenience,  the MCs with their mutual distance ($R_{LS}$)
less than 50 kpc for the last 2 Gyr are defined as a binary
(thus the models are regarded as successful).
We confirm that in the 5000 models with
$-87 \le v_{\rm x} \le 40$ (km s$^{-1}$),
$-268 \le v_{\rm y} \le -185$ (km s$^{-1}$),
and $149 \le v_{\rm z} \le 252$ (km s$^{-1}$)
for the common halos with $M_{\rm ch}\le 8 \times 10^{10} M_{\odot}$,
the number fraction of the successful models 
is 0.43. 
For comparison,  we also investigate models without common halos in which
$M_{\rm ch}$ is set to be 0.0.

In order to mimic the orbital evolution  of the MCs 
in the first passage scenario (B07),
we investigate ``the truncated halo models''
in which the density of the isothermal halo of the Galaxy
become zero beyond the truncation radius at which
the total mass becomes $10^{12} M_{\odot}$ consistent with
the observed total mass of the Galaxy (Wilkinson \& Evans 1999).
For this truncated halo model, the LMC can arrive in the Galaxy
quite recently ($\sim 0.1$ Gyr ago) for the first time.
We try to understand  how the common halo can change the orbits
of the MCs by investigating truncated halo models with and without
common halos.
In the present models,  the orbital evolution of the MCs
depends much more strongly on $M_{\rm ch}$ than on $a_{\rm ch}$
for models with and without truncation of the Galactic halo.
We thus mainly show the results of models with 
$a_{\rm ch}=10$ kpc and different
$M_{\rm ch}$.
The time $T=0$ and $T=-2.7$ Gyr correspond to the present
and the start (i.e., past) of orbital calculation,
respectively.

\subsection{Test-particle simulations}

The main purpose of this Letter is to demonstrate that
the MCs can not only keep their binary status but also 
interact strongly with the Galaxy for the last $\sim 2$ Gyr
owing to the presence of the common halo.
We however consider that it is also important to 
show that the formation of the MS is possible for some orbital models
by using idealized  test-particle simulations in which
self-gravity of particles is not included.
We therefore investigate the last 2.7 Gyr 
evolution of an exponential disk of the SMC
that has the scale-length of 2.5 kpc and is composed of 50000 particles:
the MS is here considered to be formed from the SMC as demonstrated
by GN.

For a given set of initial positions and velocities 
of the MC system at $T=-2.7$ Gyr  derived from the backward
integration scheme described in \S 2.1,
we integrate the equation of motion forward between $T=-2.7$ and
$T=0$ (i.e., the present) for the MC system
so that we can investigate dynamical evolution of the
SMC disk.
We here illustrate the results of a simulation
based on the orbital model with 
$M_{\rm S} =  3 \times 10^{9} M_{\odot}$,
$M_{\rm L} =  10^{10} M_{\odot}$,
$M_{\rm ch} =  8 \times 10^{10} M_{\odot}$,
and $a_{\rm ch} =  15$ kpc,
because this model shows the formation of the MS.
It should be stressed here that this model is used
just for an illustrative purpose: fully self-consistent
numerical simulations of the MS formation 
under the common halo scenario  will be done in our future studies.

\section{Results}

Fig. 1 shows that the MCs in the model with
the common halo can not only
keep their binary status  (i.e., $R_{\rm LS}<50$ kpc) 
for more than 2 Gyr 
but also have orbital periods of $\sim 1.5$ Gyr.
The model without the common halo, on the other hand, shows that (i) the MCs can
start their tidal interaction only quite recently ($T=-0.2$ Gyr)
and (ii) they are orbiting the Galaxy independently from each other 
for $T<-0.6$ Gyr.
These results clearly demonstrate that the common halo can play
a role in keeping the binary status of the MCs
through its gravitational influence.
The derived results of the small $R_{\rm LS}$ at $T=-1.6$ Gyr and $=-0.2$ Gyr
in the model with the common halo are strikingly similar to
those of the successful MS model by GN, which implies that
the common halo may well also play a key role in the formation
of the MS and the Magellanic Bridge.

Fig. 2 shows that the models with $M_{\rm ch}$ equal to or larger
than $4.0 \times 10^{10} M_{\odot}$ can keep the binary status of the MCs 
owing to the presence of the common halos.
The models with $2.0 \times 10^{10} M_{\odot}$ and a smaller
$a_{\rm ch}$ of 5 kpc can also keep the binary status,
though models with $M_{\rm ch}=10^{10} M_{\odot}$ can not keep the binary status
independent of $a_{\rm ch}$.
These results suggest that the common halo of the MCs needs to be
at least as  massive as  $2.0 \times 10^{10} M_{\odot}$ in order for the
MCs to keep their binary status for the last 2 Gyr.
As shown in Fig. 2, the detail of the time evolution of 
the  binary orbit ($R_{\rm ch}$) is quite different 
between the two models with $M_{\rm ch} \ge 4.0 \times 10^{10} M_{\odot}$:
formation processes of the MS, which depend on the evolution
of $R_{\rm ch}$, can be controlled by $M_{\rm ch}$.

Fig. 3 shows that the truncated halo model with the common halo
of the MCs can keep the binary status, though the orbital period
becomes significantly longer ($\sim 2.5$ Gyr) both for the LMC and for
the SMC. 
The truncated halo model without the common halo of the MCs
shows that the LMC can arrive in the Galaxy only quite recently
($T=-0.1$ Gyr) for the first time, which is consistent with
the result by B07. The SMC can interact 
strongly with the LMC at $T=-0.2$ Gyr for the first time,
which is essentially the same as the result obtained in
the model without the common halo shown in Fig. 1.
These results shown in Figs. 1 and 3 thus suggest
that the MCs can interact with each other and with the Galaxy
for the last more than 2 Gyr owing to the presence
of the common halo, even if they have the high present 3D velocities.

Fig. 4 shows that tidal streams from the SMCs  can be formed
as a result of strong interaction between the MCs and the Galaxy
for the last 2.7  Gyr.   The locations of the streams are similar
to the observed MS (e.g., Putman et al. 1998, P98) so that
the stream may well be observationally identified as the MS.
It should be however stressed that (i) the distribution of the stripped
particles in this model appears to be too dispersed in comparison
with the successful MS model by GN and  (ii) leading-arm features
observed by P98
and shown in GN can not be so clearly seen in this model.
Therefore, this model  should be regarded not as being
a successful MS model
but as suggesting a possibility that successful MS models
may well be constructed by including the common halo
in the models for reasonable values of $M_{\rm ch}$ and $a_{\rm ch}$.
It is our future work to discuss whether the observed fundamental properties
of the MS (e.g., P98) can be explained by the MS models with  the MCs 
having  common halos  based on fully self-consistent  N-body simulations
of the MS formation.

\section{Discussion and conclusions}

The present study  has first shown that if the MCs have
a common halo,  they can not only have the present
3D velocities consistent with K06 but also keep
their binary status within  the last more than 2  Gyr
in some models.
It should be however stressed that the common diffuse halo needs to
have (i) the present velocity ($|v|$) significantly smaller than
that of the LMC  and (ii) the mass larger than
$\sim 2 \times 10^{10}$ in order for the MCs to keep their binary
status for the last $\sim 2$ Gyr.
These two  requirements would be  very hard to be confirmed
directly by observations:
differences in velocities between the center of the common
halo and  those of the MCs and the mass of the halo
can be inferred from numerical simulations on the binary formation
of the MCs.
Then, how could  they have formed  a common halo in the histories of the MCs  ?

We here suggest the following two scenarios for the common halo formation.
The first is that the MCs might have dynamically coupled recently 
($<4$ Gyr)
to form a common halo: 
dynamical relaxation processes  of 
the two pre-existing halos of the MCs
during binary galaxy formation can be  responsible for the common halo
formation.
The orbital evolution models including dynamical friction between
the MCs
by BC05 showed  that the MCs could  become
dynamically coupled for the first time about $3-4$ Gyr ago.
Recent cosmological N-body simulations of the pair galaxy formation
based on a  $\Lambda$ cold dark matter ($\Lambda$CDM) cosmology 
have shown that the pair formation like the MCs can occur
at $z<0.33$ corresponding to less than 3.7 Gyr ago for a canonical
set of cosmological parameters (Ishiyama et al. 2008).
These results imply that the common halo formation of the MCs 
might have happened recently ($<4$ Gyr ago).
We suggest that the required larger  mass of the common halo
(i.e., $M_{\rm ch} \ge 2\times 10^{10} M_{\odot}$) for the binary
statue of the MCs in the last $\sim 2$ Gyr
can come from the stripped dark matter halos of the MCs at the epoch of 
their binary formation: their original masses are significantly
larger than the present ones.

The second scenario  is that 
a small group of galaxies including the MCs 
in its central region fell onto the outer region of the Galaxy
and then lost most of the halo and the group member galaxies
via tidal stripping by the Galaxy: the MC system  with the common
halo is the remnant of a destroyed group.
Li \& Helmi (2008) have investigated merging histories of
subhalos in a Milky Way-like halo using
high-resolution simulations based on  a $\Lambda$CDM model
and thereby demonstrated  that about one-third of the subhalos
have been accreted in groups (see also Sales et al. 2007 and  Ludlow
et al. 2008).
The demonstrated higher incidence of the group infall 
appears to suggest that the MCs can  originate from
a group thus that the second scenario is also viable.
It is currently unclear which of the two scenarios are more
consistent with other observations.

If the MCs really have a  common halo,
then the common halo would have the following possible dynamical
effects on the Galaxy and the LMC.
Firstly, the MC system embedded in the common halo, which is more massive than
the LMC,  can more strongly influence the outer part
of the Galaxy 
than the LMC alone 
is demonstrated to be able to do
(e.g., Tsuchiya 2002)  so that the observed HI warp of the Galaxy
(e.g., Diplas \& Savage 1991) can be more naturally
explained in terms of the common halo scenario.
Secondly, the common halo can weakly influence the disk of the LMC
so that the combined tidal effect of the Galaxy, the SMC,
and the common halo could  form a off-center bar that is
more pronounced  
than the simulated one
in the  last LMC-SMC interaction about 0.2 Gyr ago (Bekki \& Chiba 2007). 
Thirdly, the common halo enables the LMC and the SMC to
have their  stellar halos extended much beyond their
optical radii. 

The present study suggests  that it would be  difficult to
observationally determine the 3D velocities of
the MC system solely  from proper motion measurements 
of the MCs in the common halo scenario.
Furthermore, it would be  even more difficult for 
theoretical and numerical works  to predict 
precisely the long-term orbital evolution of the MCs 
with a common halo owing to additional
two (or more) parameters for physical properties of the  halo.
If the observed proper motions of the MCs (K06) 
are really true ones of  {\it the centers of mass for the LMC
and  the SMC},  some new physical processes need to be
incorporated into the tidal models of the MS formation 
for self-consistency.
The hypothesized  common halo of the MCs in the
present study  would be  just one of possible
physical ingredients that need to be considered in constructing
a more self-consistent model of the MS formation. 

\acknowledgments
I (K.B.) am   grateful to the anonymous referee for valuable comments,
which contribute to improve the present paper.
K.B. acknowledges the financial support of the Australian Research
Council throughout the course of this work.


\newpage

\begin{deluxetable}{ccccc}
\footnotesize
\tablecaption{The present locations,  velocities, and masses  of the
LMC, the SMC, and the common halo adopted in the present
orbital evolution models.\tablenotemark{a} \label{tbl-1}}
\tablewidth{0pt}
\tablehead{
\colhead{  Component} & 
\colhead{  $(x,y,z)$} (kpc) & 
\colhead{  $(v_{\rm x},v_{\rm y},v_{\rm z})$ (km s$^{-1}$)} & 
\colhead{  Mass ($\times 10^{10} M_{\odot}$)} & 
\colhead{  Effective radius (kpc)}  }
\startdata
LMC & (-1,-41,-27) &  (-86,-268,252) & 1-2 & 3    \\
SMC &  (14,-34,-40) & (-87,-247,149) & 0.3-0.6 & 2    \\
Common halo &  (1, -40, -28)  &  (1,-220,191) &  1-8 & 5-15    \\
\enddata
\tablenotetext{a}{The parameter ranges investigated
in the present study are shown for  $M_{\rm L}$, $M_{\rm S}$,
$M_{\rm ch}$, and $a_{\rm ch}$.}
\end{deluxetable}


\newpage
\begin{figure}
\plotone{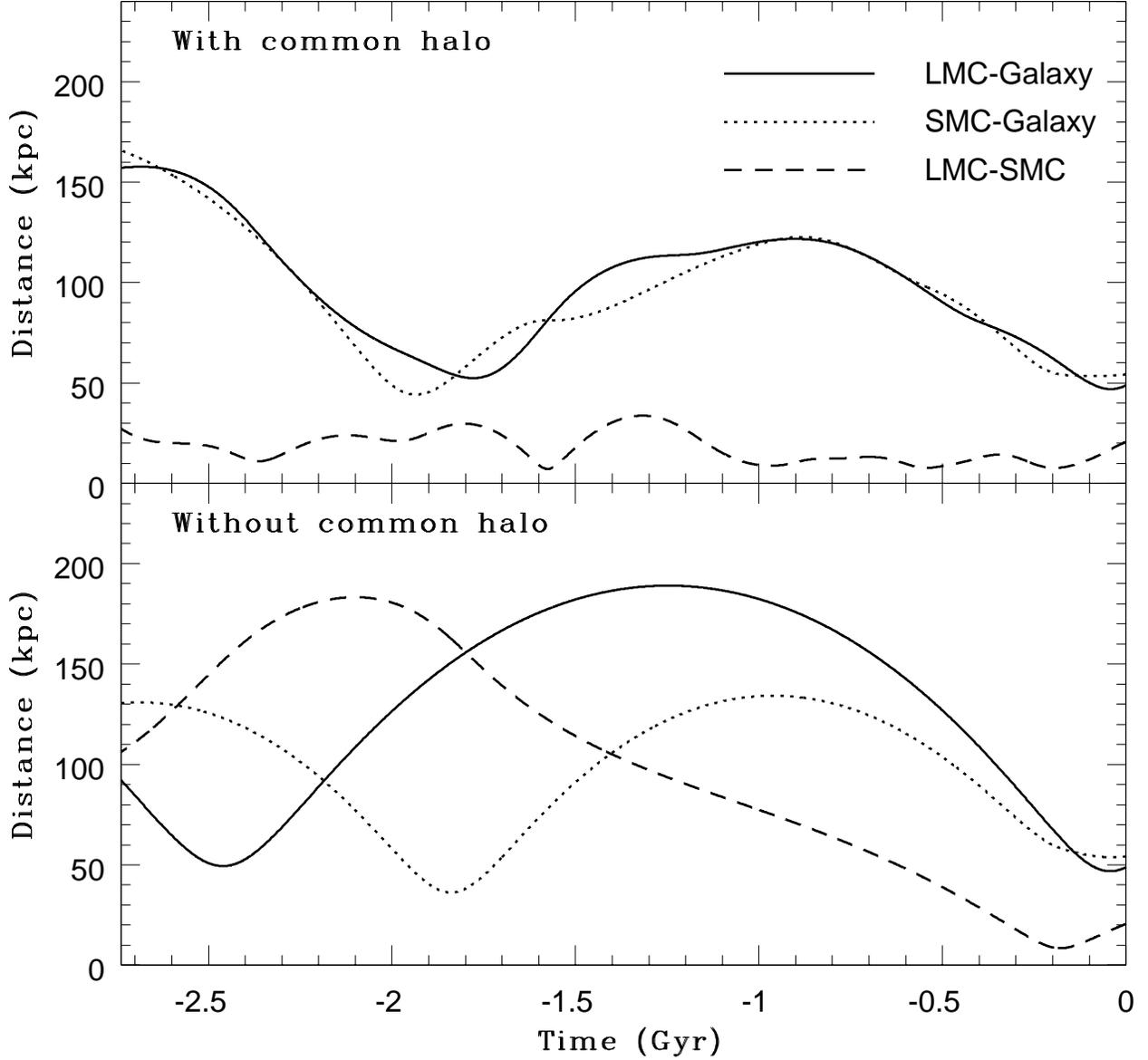}
\figcaption{
Time evolution of the distances between the LMC and the Galaxy (solid),
the SMC and the Galaxy (dotted), and the LMC and the SMC (dashed)
for the last i$\sim 2.7$  Gyr in the model with (upper) and
without (lower) the common halo of the MCs.
The time $T=0$ Gyr and $T=-2.7$ Gyr mean the present and 2.7 Gyr ago,
respectively,
in this figure. 
For these models,  $M_{\rm L} =2.0  \times 10^{10} M_{\odot}$,
$M_{\rm S} =3.0  \times 10^{9} M_{\odot}$,
$M_{\rm ch} =4.0  \times 10^{10} M_{\odot}$,  and
$a_{\rm ch}=10$ kpc are assumed.
Note that the MCs can keep their binary status for the last 2.7 Gyr
only for the model with the common halo.
\label{fig-1}}
\end{figure}

\begin{figure}
\plotone{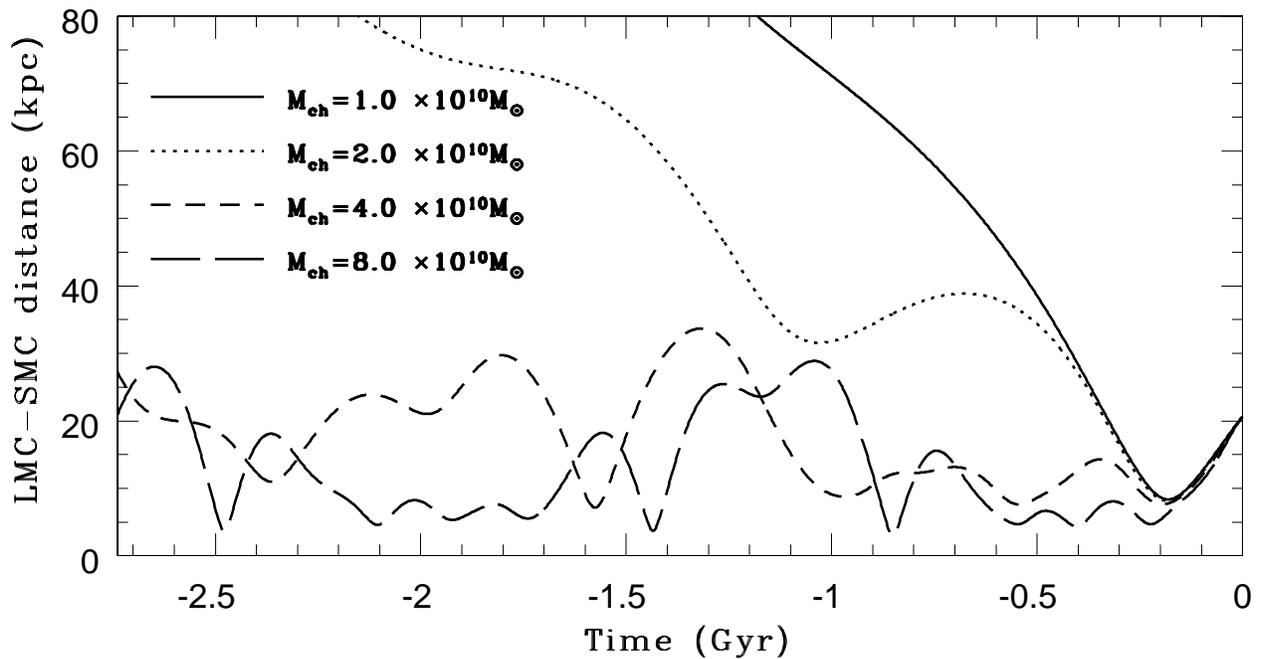}
\figcaption{
Time evolution of the distances between the LMC and the SMC  ($R_{\rm LS}$)
for the last $\sim 2.7$ Gyr
in the models with $M_{\rm ch} =1.0  \times 10^{10} M_{\odot}$  (solid),
$M_{\rm ch} =2.0  \times 10^{10} M_{\odot}$  (dotted),
$M_{\rm ch} =4.0  \times 10^{10} M_{\odot}$  (short-dashed),
and $M_{\rm ch} =8.0  \times 10^{10} M_{\odot}$  (long-dashed).
For these models,  $M_{\rm L} =2.0  \times 10^{10} M_{\odot}$,
$M_{\rm S} =3.0  \times 10^{9} M_{\odot}$,
and $a_{\rm ch}=10$ kpc are assumed.
\label{fig-2}}
\end{figure}

\begin{figure}
\plotone{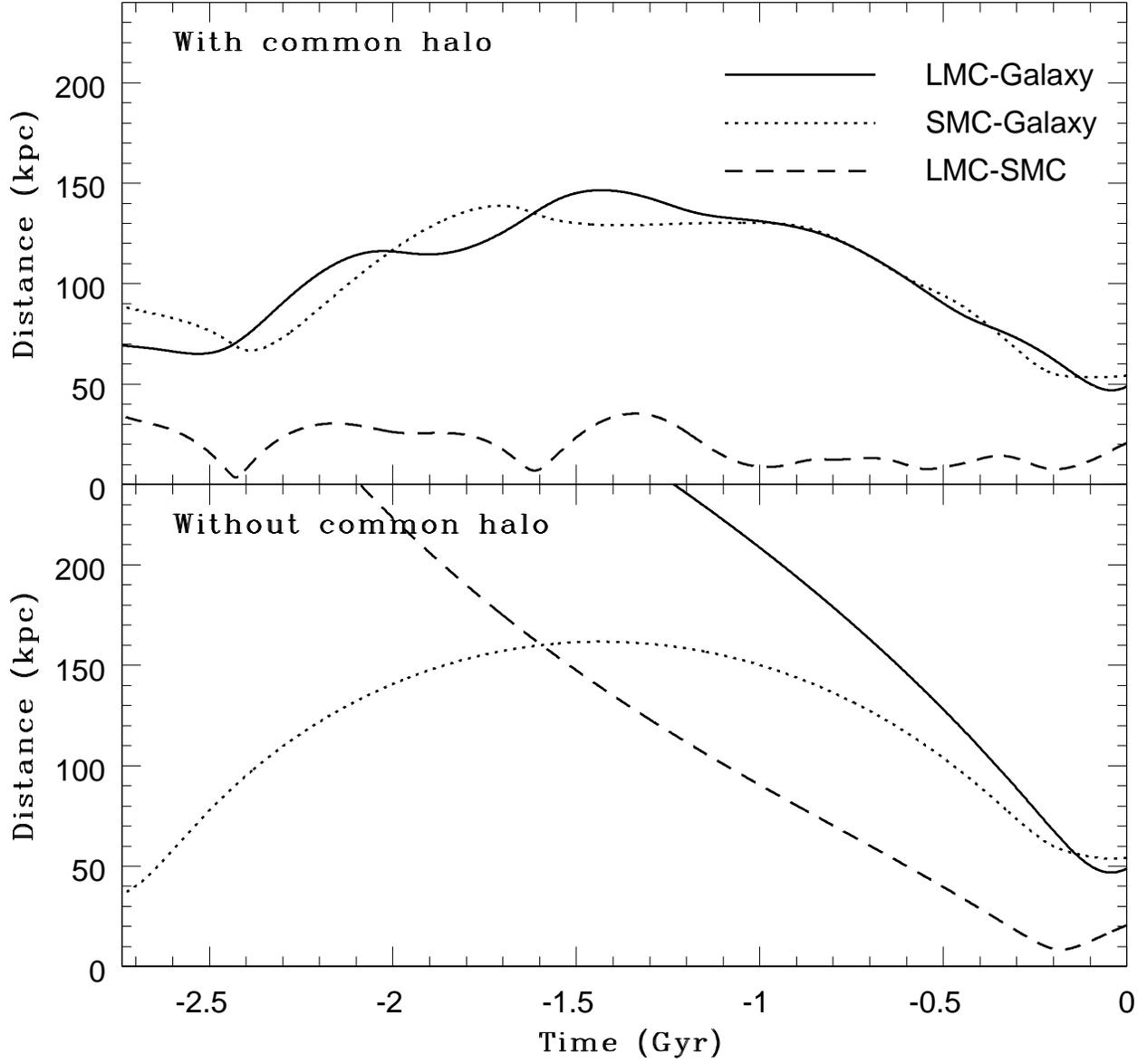}
\figcaption{
The same as Fig.1 but for the truncated halo models. 
\label{fig-3}}
\end{figure}

\begin{figure}
\epsscale{.70}
\plotone{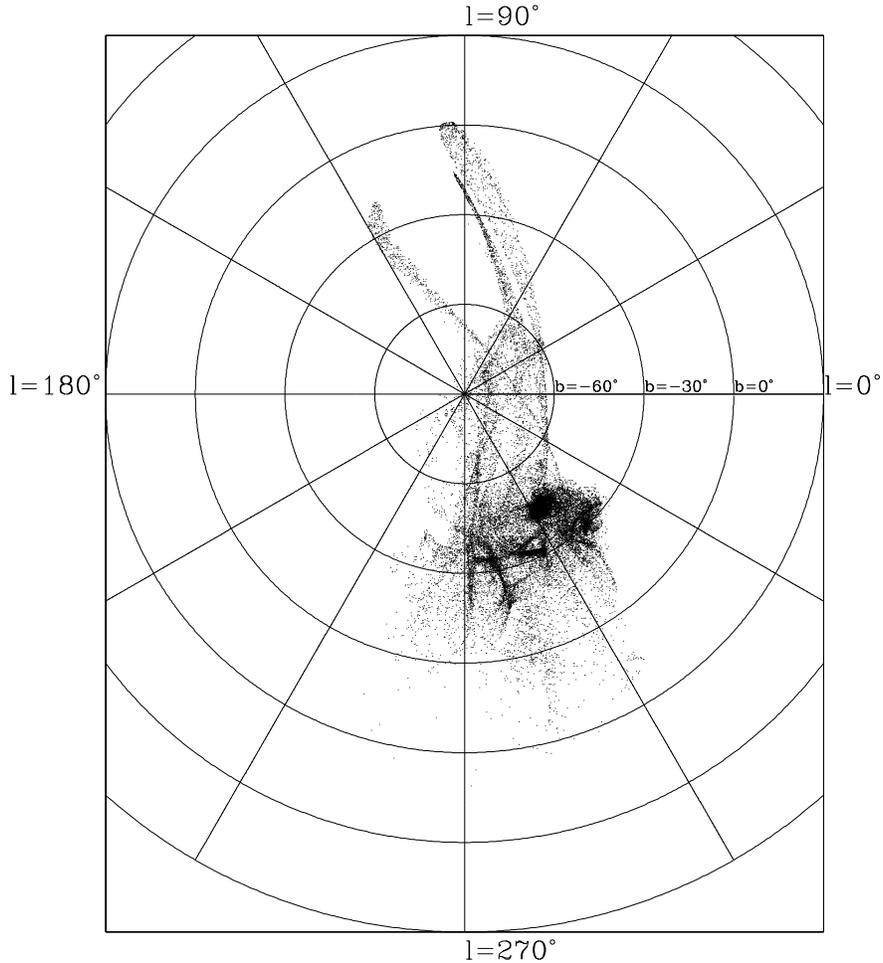}
\figcaption{
The present distribution of particles of the SMC
projected onto  the Galactic coordinate $(l,b)$ in 
the test-particle simulation 
based on the orbital model with
$M_{\rm L} =1.0  \times 10^{10} M_{\odot}$,
$M_{\rm S} =3.0  \times 10^{9} M_{\odot}$,
$M_{\rm cl}=8.0 \times 10^{10} M_{\odot}$,
and $a_{\rm cl}=15$ kpc. The particles stripped from the SMC
as a result of the LMC-SMC-Galaxy interaction
can form streams and the locations of the streams appear to be similar
to the observed location of the MS (e.g., P98).
This result implies  a possibility that successful MS models
can be constructed 
by our future
self-consistent N-body simulations
in the common halo scenario.
\label{fig-4}}
\end{figure}

\end{document}